# Image based Cryptography from a distance

Yousuf Ibrahim Khan, Saad Mahmud Sonyy, S.M. Musfequr Rahman

**Abstract**— An information is a message which is received and understood. Information can be sent one person to another over a long range but the process of sending information must be done in a secure way especially in case of a private message. Mathematicians and Engineers have historically relied on different algorithmic techniques to secure messages and signals. Cryptography, to most people, is concerned with keeping communications private. Indeed, the protection of sensitive communications has been the emphasis of cryptography throughout much of its history. Sometimes it is safer to send a message using an image and thus cryptography can also be done using images during an emergency. The need to extract information from images and interpret their contents has been one of the driving factors in the development of image processing and cryptography during the past decades. In this paper, a simple cryptographic method was used to decode a message which was in an image and it was done using a popular computational software.

—————————— ◆ ——————————

## 1 INTRODUCTION

Encoding is the transformation of data into some unreadable form. Its purpose is to ensure privacy by keeping the information hidden from anyone for whom it is not intended. Decoding is the reverse of encoding ; it is the transformation of encrypted data back into some intelligible form. Cryptography is popularly known as the study of encoding and decoding private massages. Because its importance, there is a recent surge of interest in cryptography. In this subject, codes are called ciphers, uncoded massages are called plaintext, and coded messages are called ciphertext. The method of converting from plaintext to ciphertext is called enciphering, and the method of converting from ciphertext to plaintext is called deciphering. In this research, an image was used where a hidden message was written.

This message or code was encrypted using Hill Cipher technique (one the cryptographic methods). Artificial Neural Networks are very good in learning from experience. If characters are used to train the network then in future the network will also be able to identify one of those characters or alphabets. Using some basic image processing techniques and artificial neural network one can identify the characters or some patterns in character arrangement from the image and decrypt it using a computer program. The image could also be taken in Real-Time. In this paper techniques to acquire images in Real-Time using a webcam were also shown. This method could become useful in security and military applications where text messages and conversations are not safe.

## 2 ANN IN CHARACTER RECOGNITION

Significant progress in the development of machine vision and image processing technology has been made in the past few years in conjunction with improvements in computer technology [1]. It is presently quite difficult to use machine or computer vision to distinguish, recognize characters from the image in real time, due to the substantial computational resources and the complicated algorithms required. Artificial neural networks (ANNs) can overcome some of these difficulties by interpreting images quickly and effectively. ANNs are composed of numerous processing elements arranged in various layers, with interconnections between pairs of processing elements [2][3][4]. They are designed to emulate the structure of natural neural networks such as those of a human brain.

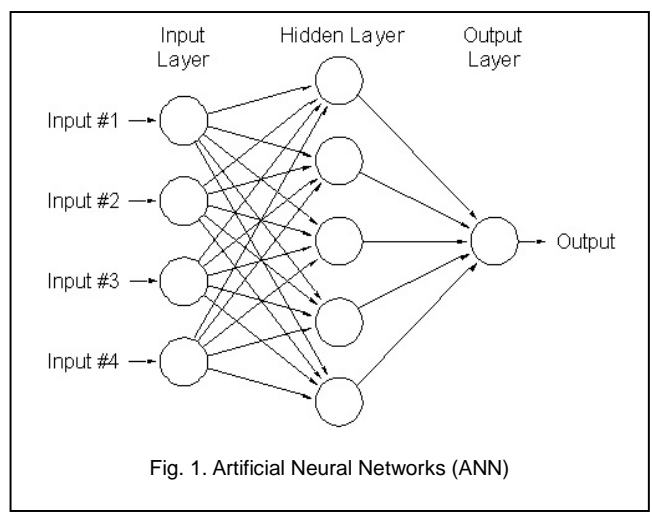

Fig. 1. Artificial Neural Networks (ANN)

So an Artificial Neural Network (ANN) is a mathematical or computational model based on the structure and functional aspects of biological neural networks [5]. It consists of an interconnected group of artificial neurons, and it processes information using a connectionist approach.

An ANN mostly is an adaptive system that changes its structure based on external or internal information that flows through the network during the learning phase. The main feature of an ANN is its ability to learn complex functional relations by generalizing from a limited amount of training data. Neural nets can thus be used as (black-box) models of nonlinear, multivariable static and dynamic systems and can be trained by using input–output data observed on the system.

## 2.1 How it Learns ?

A typical feed forward neural network has an input, a hidden and an output layer which is shown in Figure. 1. Each component includes a neuron, weights and a transfer function. An input $X_j$ is transmitted through a connection which multiplies its strength by a weight $W_{ij}$ to give a product $X_jW_{ij}$. The product is an argument to a transfer function F which yields an output $Y_i$ represented by:

$$Y_i = F\left(\sum_{j=1}^{n} X_j W_{ij}\right) \quad (1)$$

Where i is a neuron index in the hidden layer and j is an index of an input to the neural network. Training is the process of modifying the connection weights in some orderly fashion using a suitable learning method. So applying artificial neural networks can be valuable in English alphabets.

## 2.2 Character Recognition

Character recognition, usually abbreviated to optical character recognition or shortened OCR, is the mechanical or electronic translation of images of handwritten, typewritten or printed text into machine-editable text. It is a field of research in pattern recognition, artificial intelligence and machine vision. Though academic research in the field continues, the focus on character recognition has shifted to implementation of proven techniques.

OCR software can recognize a wide variety of fonts, but handwriting and script fonts that mimic handwriting are still problematic, therefore additional help of neural network power is required. Developers are taking different approaches to improve script and handwriting recognition [6]. As mentioned above, one possible approach of handwriting recognition is with the use of neural networks. In this paper Feed-Forward Neural network was chosen for simulate recognition problem.

## 2.3 Application of Image Processing

The process which was shown in this paper was also an application of Image Processing. So before going further into discussion it is necessary to give some basics of image processing.

An image is a visual representation of an object, a person, or a scene produced by an optical device such as a mirror, a lens, or a camera, web cam. This representation is two dimensional (2D), although it corresponds to one of the infinitely many projections of a real-world, three-dimensional (3D) object or scene [7]. In image processing various techniques and algorithms are used to perform tasks like image enhancement, noise removal, color detection, edge detection, deblurring, sharpening etc. Some of the applications are given below :

**Medical Applications:** Diagnostic imaging modalities such as digital radiography, PET (positron emission tomography), CAT (computerized axial tomography), MRI (magnetic resonance imaging), and fMRI (functional magnetic resonance imaging), among others, have been adopted by the medical community on a large scale.

**Industrial Applications:** Image processing systems have been successfully used inmanufacturing systems formany tasks, such as safety systems, quality control, and control of automated guided vehicles (AGVs).

**Military Applications:** Some of the most challenging and performance-critical scenarios for image processing solutions have been developed formilitary needs, ranging from detection of soldiers or vehicles to missile guidance and object recognition and reconnaissance tasks using unmanned aerial vehicles (UAVs). In addition, military applications often require the use of different imaging sensors, such as range cameras and thermographic forward-looking infrared (FLIR) cameras.

**Law Enforcement and Security:** Surveillance applications have become one of the most intensely researched areas within the video processing community. Biometric techniques (e.g., fingerprint, face, iris, and hand recognition), which have been the subject of image processing research for more than a decade, have recently become commercially available.

**Consumer Electronics:** Digital cameras and camcorders, with sophisticated built-in processing capabilities, have rendered film and analog tape technologies obsolete. Software packages to enhance, edit, organize, and publish images and videos have grown in sophistication while keeping a user-friendly interface.High-definition TVs,monitors,DVDplayers, and personal video recorders (PVRs) are becoming increasingly popular and affordable. Image and video have also successfully made the leap to other devices, such as personal digital assistants (PDAs), cell phones, and portable music (MP3) players.

## 3 IMAGE ACQUISITION AND PROCESSING

The problem which has been solved in this research paper doesn't accound image acquisition in Real-Time. But the image can be taken using Real-Time which is more state of the art. Here cryptography is shown using a captured image which was created using a word processor but Real-Time image processing techniques are shown below in brief :

To start with working in real time, one must have a functional USB webcam or an acquisition device connected to PC and its driver installed. MATLAB has built-in adaptors for accessing these devices. An adaptor is a software that MATLAB uses to communicate with an image acquisition device.

One can check if the support is available for his camera in MATLAB by typing the following which gives Figure. 2. :

\>> imaqhwinfo

% stands for image acquisition hardware info

\>> cam=imaqhwinfo;

\>> cam.InstalledAdaptors

```
ans =

    InstalledAdaptors: {'coreco'   'winvideo'}
         MATLABVersion: '7.5 (R2007b)'
          ToolboxName: 'Image Acquisition Toolbox'
       ToolboxVersion: '3.0 (R2007b)'

ans =

    'coreco'    'winvideo'
```
Fig. 2. This was the output of the computer used in this research and it will be slightly different on another computer

To get more information about the device, type

\>> dev_info = imaqhwinfo('winvideo',1)

This will give the output of Figure. 3

```
dev_info =

             DefaultFormat: 'YUY2_160x120'
         DeviceFileSupported: 0
                  DeviceName: 'Vimicro USB Camera (Altair)'
                    DeviceID: 1
           ObjectConstructor: 'videoinput('winvideo', 1)'
            SupportedFormats: {1x5 cell}
```
Fig. 3. Device Information

\>>dev_info.SupportedFormats %which gives Figure. 4

Now one can preview the video captured by the camera by defining an object (say by the name 'vid') and associate it with the webcam (An example of Webcam is shown in Figure. 5).

\>>vid=videoinput('winvideo',1, 'YUY2_160x120');
\>>preview(vid) % Gives the video output like Figure. 6

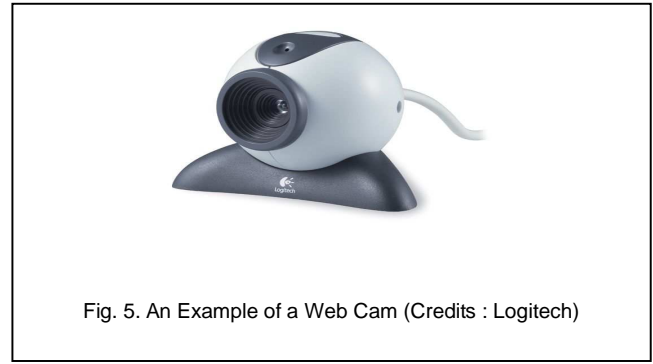
Fig. 4. Supported video size formats

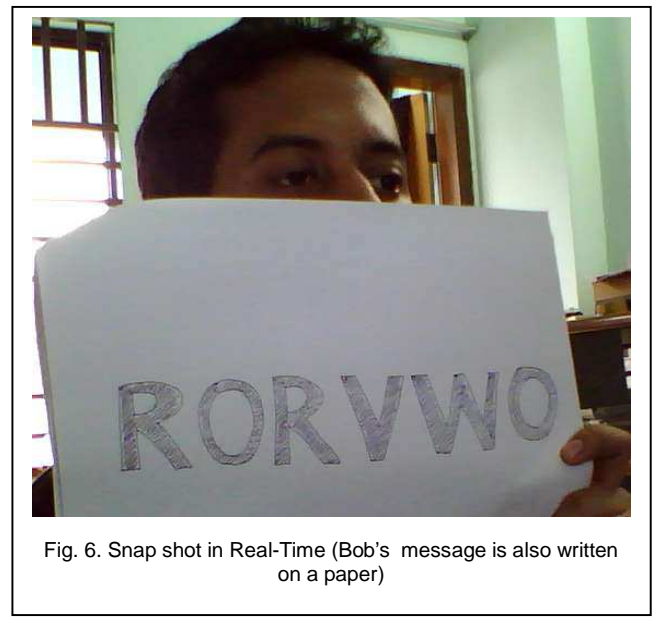
Fig. 5. An Example of a Web Cam (Credits : Logitech)

Fig. 6. Snap shot in Real-Time (Bob's message is also written on a paper)

The following codes captured the image :

\>>im=getsnapshot(vid);    % where vid is video input object
\>>imshow(im);
\>>im=ycbcr2rgb(im);
\>>imshow(im);
\>>imwrite(im,'myimage.jpg');

## 4 CRYPTOGRAPHY

Due to the rapid growth of digital communication and electronic data exchange, information security has become a crucial issue in industry, business, and administration. Modern Cryptography provides essential techniques for securing information and protecting data. The rapid growth of electronic communication means that issues in information security are of increasing practical importance. Messages exchanged over worldwide publicly accessible computer networks must be kept confidential and protected against manipulation.

Applications like Electronic business requires digital signatures that are valid in law, and secure payment protocols. Modern cryptography provides solutions to all these problems. So the study of encoding and decoding secret messages is called Cryptography. Although secret codes date to the earliest days of written communication, there has been a recent surge of interest in the subject because of the need to maintain the privacy of information over public lines of communication [8][9].

There are several cryptography techniques available but for simplicity Hill Cipher method was used in this paper.

### 4.1 Hill Cipher Technique

This technique was named after Lester. S. Hill, who introduced this in his two papers [10][11]. Hill cipher is the idea of encoding plaintext group by group, not just letter by letter. To get an idea suppose that each plaintext and ciphertext letter except Z is assigned the numerical value that specifies its position in the standard alphabet (Table 1). For reasons that will become clear later sections, Z is assigned a value of zero.

TABLE 1
ALPHANUMERICS FOR HILL CIPHER

| A | E | I | M | Q | U | Y |
|---|---|---|---|---|---|---|
| 1 | 5 | 9 | 13 | 17 | 21 | 25 |
| B | F | J | N | R | V | Z |
| 2 | 6 | 10 | 14 | 18 | 22 | 0 |
| C | G | K | O | S | W |   |
| 3 | 7 | 11 | 15 | 19 | 23 |   |
| D | H | L | P | T | X |   |
| 4 | 8 | 12 | 16 | 20 | 24 |   |

### 4.2 Modular Arithmatic and Hill 2-cipher

In this section the cryptography technique that used in the paper has been discussed in mathematical term. In the simplest hill ciphers, pairs of plaintext are transformed into ciphertext by the method below:

Step 1. A 2X2 matrix is chosen with integer entries

$$A = \begin{bmatrix} a_{11} & a_{12} \\ a_{21} & a_{22} \end{bmatrix} \quad (2)$$

to perform the encoding. Certain additional conditions on A will be imposed later.

Step 2. Successive plaintext letters are grouped into pairs, an arbitrary "dummy" letter is added to fill out the last pair if the plaintext has an odd number of letters, and each plaintext letter is replaced by its neumerical value.

Step 3. Each plaintext pair $p_1p_2$ is converted into a column vector and the product Ap is formed. P is called a plaintext vector and Ap the corresponding ciphertext vector.

$$p = \begin{bmatrix} p_1 \\ p_2 \end{bmatrix} \quad (3)$$

Step 4. Each ciphertext vector is converted into its alphabetic equivalent.

At first suppose that there are two persons in the event. The first person Bob is sending a message using hill cipher and he wrote that message on a paper. Think that Bob is using a webcam in case of an emergency. Here we showed similar scenario in a bitmap(.bmp) (image with dots) image (Fig. 7.)

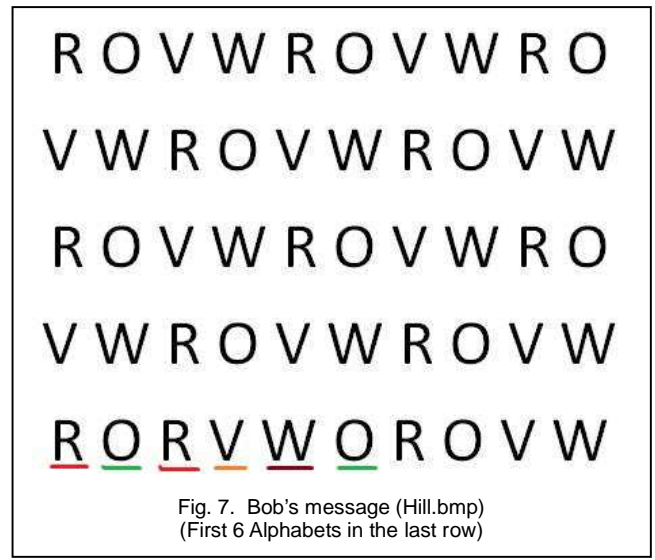

Fig. 7. Bob's message (Hill.bmp)
(First 6 Alphabets in the last row)

Suppose Bob was in a danger and he wanted to say 'HELP ME' but he didn't write it like that. Instead he transferred his message using Hill – 2 cipher technique.

He did that using following steps :

He grouped the plaintext into pairs and obtained

**HE LP ME**

The original words had even numbers of letters in it. If it were odd then a dummy letter was required to make it even. From table 1. it becomes

$$8 \quad 5 \quad\quad 12 \quad 16 \quad\quad 13 \quad 5$$

Now suppose he used an arbitrary 2X2 matrix [ 1 2 ; 0 3]; To encipher the pair HE, it becomes

$$\begin{bmatrix} 1 & 2 \\ 0 & 3 \end{bmatrix} \begin{bmatrix} 8 \\ 5 \end{bmatrix} = \begin{bmatrix} 18 \\ 15 \end{bmatrix} \quad\quad (4)$$

Which from table 1, yields the ciphertext **RO**

To cipher LP, it becomes

$$\begin{bmatrix} 1 & 2 \\ 0 & 3 \end{bmatrix} \begin{bmatrix} 12 \\ 16 \end{bmatrix} = \begin{bmatrix} 44 \\ 48 \end{bmatrix} = \begin{bmatrix} 18 \\ 22 \end{bmatrix} \quad\quad (5)$$

However, there is a problem there, the number 44 had no alphabet equivalent (Table 1). To resolve Bob used modular arithmetics :

"Whenever an integer greater than 25 occurs, it will be replaced by the remainder that results when this integer is divided by 26".

Because the remainder after division by 26 is one of the integers 0, 1, 2, …. 25, this procedure will always yield an integer with an alphabet equivalent. Thus, he replaced 44 by 18, which is the remainder after dividing 44 by 26. So from table 1. The pair LP becomes RV.

And the pair ME becomes WO,

$$\begin{bmatrix} 1 & 2 \\ 0 & 3 \end{bmatrix} \begin{bmatrix} 13 \\ 5 \end{bmatrix} = \begin{bmatrix} 23 \\ 15 \end{bmatrix} \quad\quad (6)$$

So without spaces his message 'HELPME' becomes '**RORVWO**'. Because the plaintext was grouped in pairs and enciphered by a 2X2 matrix, this Hill Cipher is referred to as a Hill 2-Cipher.

In general, for a Hill n-Cipher, plaintext is grouped into sets of n letters and enciphered by an nxn matrix.

Suppose his friend Albert knew he would use Hill 2-Cipher technique, so Albert decoded the message like this :

In modular arithmetic, if a and m have no common prime factors, then a has a unique reciprocal modulo m; conversely, if a and m have a common prime factor, then a has no reciprocal modulo m.

As an example the number 4 has no reciprocal modulo 26, because 4 and 26 have a common prime factor (Table 2).

Now in case of Hill cipher, decipherment uses the inverse (mod 26) of the enciphering matrix. To be precise, if m is a positive integer, then a square matrix A with entries in $Z_m$ is said to be invertible modulo m if there is a matrix B with entries in $Z_m$ such that

$$AB = BA = I \; (\text{mod } m)$$

TABLE 2
RECIPROCALS MODULO 26

| a | 1 | 3 | 5 | 7 | 9 | 11 |
|---|---|---|---|---|---|---|
| $a^{-1}$ | 1 | 9 | 21 | 15 | 3 | 19 |
| a | 15 | 17 | 19 | 21 | 23 | 25 |
| $a^{-1}$ | 7 | 23 | 11 | 5 | 17 | 25 |

Suppose now that A the matrix which was used by Bob was invertible modulo 26 and Albert knew this exact arbitrary matrix (Let it was decided earlier between them) then Bob's plaintext vector was c = Ap (mod 26).

Now the reverse process is p = $A^{-1}$ c (mod 26)

Thus, Albert can recover his plaintext vector from the corresponding ciphertext vector by multiplying it on the left by $A^{-1}$ (mod 26). In cryptography it is important to know which matrices are invertible modulo 26 and how to obtain their inverses. In ordinary arithmetic, a square matrix A is invertible if and only if det (A) ≠ 0, or equivalently, if and only if det (A) has a reciprocal. So a square matrix A with entries in $Z_m$ is invertible modulo m if and only if the residue of det (A) modulo m has a reciprocal modulo m. Also m and the residue of det (A) modulo m have no common prime factors.

So if A is like,

$$A = \begin{bmatrix} a & b \\ c & d \end{bmatrix}$$

And has entries in $Z_{26}$ and the residue of det(A) = ad-bc modulo 26 is not divisible by 2 or 13, then the inverse of A(mod 26) is given by,

$$A^{-1} = (ad\text{-}bc)^{-1} \begin{bmatrix} d & -b \\ -c & a \end{bmatrix} \quad\quad (\text{mod } 26)$$

Where (ad-bc)$^{-1}$ is the reciprocal of the residue of ad-bc (mod26).

The number 3 has a reciprocal modulo 26 because 3 and 26 have no common prime factors. This reciprocal can be ob-tained by finding the number x in $Z_{26}$ that satisfies the modular equation, $3x = 1 \pmod{26}$. From the theory of modular arithmetic if a is a number in $Z_m$, then a number a-1 in $Z_m$ is called a reciprocal of a modulo m if $aa-1 = a-1a = 1 \pmod{m}$.
So $3 \times 9 = 27 = 1 \pmod{26}$, dividing 27 by 26 the remainder is 1. The complete list is given in Table 2.

So Albert finds, $(ad-bc)^{-1} = 3^{-1} = 9$ (from Table 2)

$$A^{-1} = 9 \begin{bmatrix} 3 & -2 \\ 0 & 1 \end{bmatrix} \pmod{26}$$

$$= \begin{bmatrix} 27 & -18 \\ 0 & 9 \end{bmatrix} \pmod{26}$$

$$= \begin{bmatrix} 1 & 8 \\ 0 & 9 \end{bmatrix}$$

The (-) of -18 doesn't count in modular arithmetic so it becomes 8 by dividing 26 by 18.

If Albert multiplies,

$$\begin{bmatrix} 1 & 8 \\ 0 & 9 \end{bmatrix}\begin{bmatrix} 18 \\ 15 \end{bmatrix} = \begin{bmatrix} 138 \\ 135 \end{bmatrix} \pmod{26} = \begin{bmatrix} 8 \\ 5 \end{bmatrix}$$

$$\begin{bmatrix} 1 & 8 \\ 0 & 9 \end{bmatrix}\begin{bmatrix} 18 \\ 22 \end{bmatrix} = \begin{bmatrix} 12 \\ 16 \end{bmatrix}$$

$$\begin{bmatrix} 1 & 8 \\ 0 & 9 \end{bmatrix}\begin{bmatrix} 23 \\ 15 \end{bmatrix} = \begin{bmatrix} 13 \\ 5 \end{bmatrix}$$

And the alphabet equivalent of these vectors are,

HE    LP    ME

Which yields the message 'HELP ME'. In the next section the computer aided solution has been given using MATLAB and a Hardware (Webcam). In case of an emergency one can easily use webcam's to send his or her message and the receiver can detect the message using these described techniques.

## 5 DESCRIPTION OF THE METHOD

To solve the defined character recognition problem MATLAB (R2007b) computation software was used with Neural Network and Image Processing Toolbox. The computation steps were divided into the following categories:

**Automatic Image Processing :** The image was first converted to grayscale image followed by the threshing technique, which make the image become binary image. The binary image is then sent through connectivity test in order to check for the maximum connected component, which is, the box of the form. After locating the box, the individual characters are then cropped into different sub images that are the raw data for the feature extraction routine.

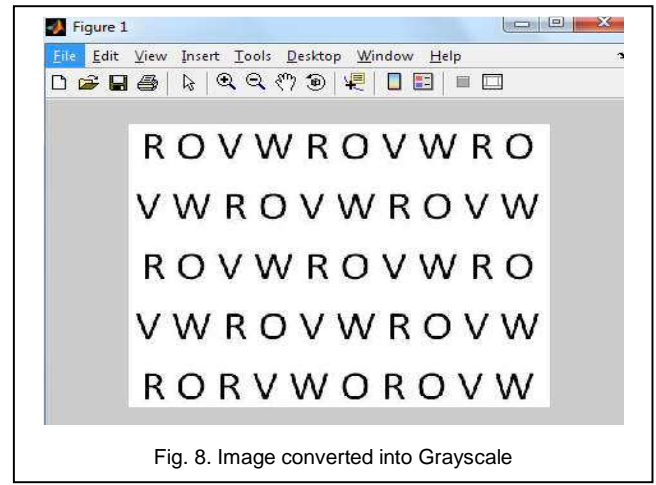

Fig. 8. Image converted into Grayscale

**Source code :**

I = imread('Hill.bmp'); % Reads the image in the directory

imshow(I); % Shows the original image

Igray = rgb2gray(I); %Grayscale conversion

imshow(Igray) % Shows the Grayscale image (Fig. 8.)

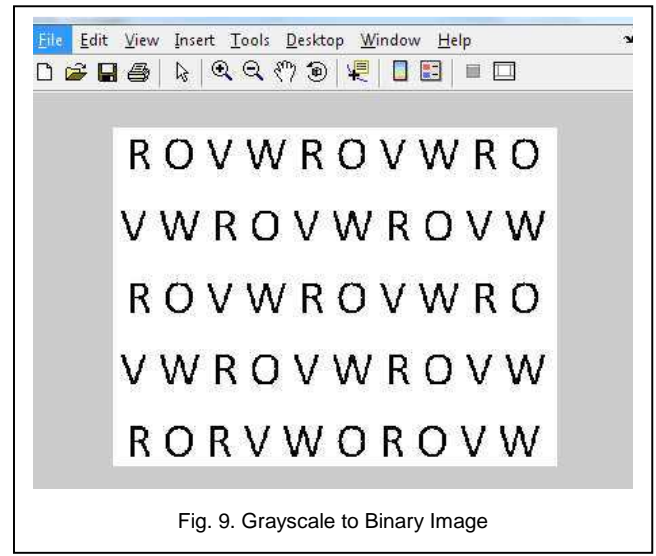

Fig. 9. Grayscale to Binary Image

Ibw = im2bw(Igray,graythresh(Igray));

%Grayscale to Binary (Fig. 9)

imshow(Ibw)

Iedge = edge(uint8(Ibw));

imshow(Iedge) % Edge Detections (Fig 10)

```
se = strel('square',2); % Image Dilation

Iedge2 = imdilate(Iedge, se);

imshow(Iedge2);

Ifill= imfill(Iedge2,'holes'); %Image Filling (Fig 11)

imshow(Ifill)

[Ilabel num] = bwlabel(Ifill); %Blobs Analysis

disp(num);

Iprops = regionprops(Ilabel);

Ibox = [Iprops.BoundingBox];

Ibox = reshape(Ibox,[4 num]);

imshow(I)

hold on;

for cnt = 1:num

rectangle('position', Ibox(:,cnt),'edgecolor','r');

%Finding object Location (Fig. 12)

end
```

These source code give the number 50, which means there are 50 objects in the original bitmap Hill.bmp image.

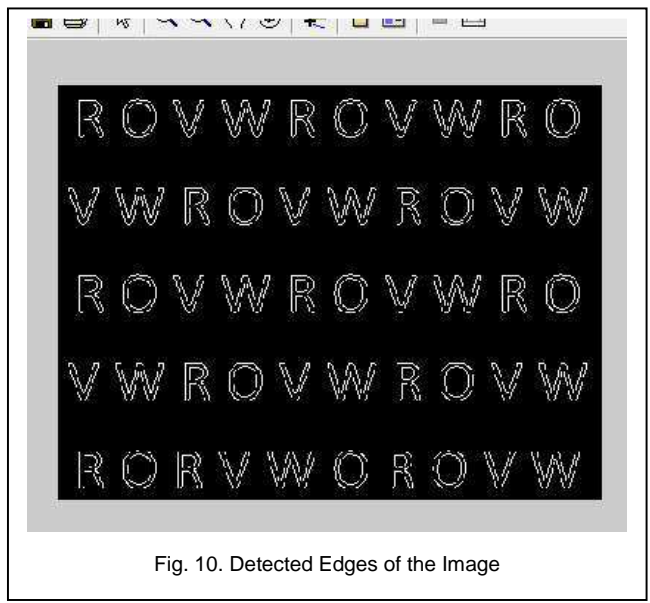

Fig. 10. Detected Edges of the Image

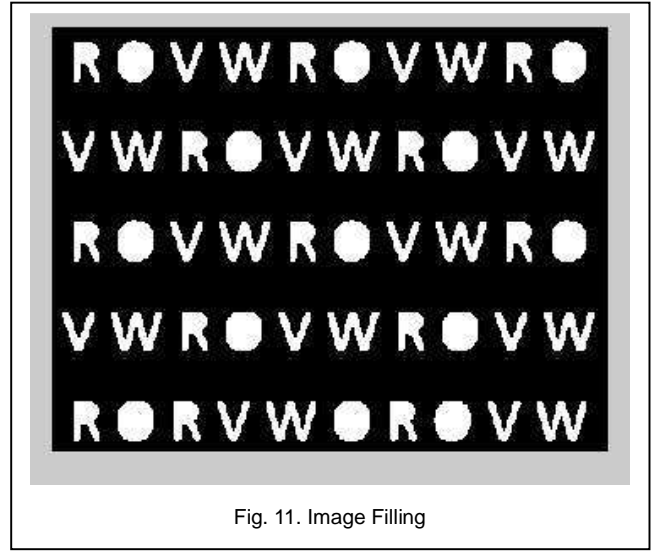

Fig. 11. Image Filling

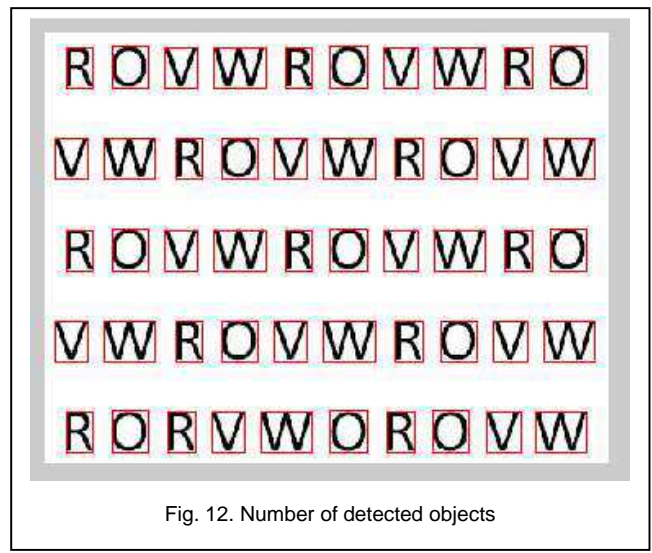

Fig. 12. Number of detected objects

**Feature Extraction :** The sub-images have to be cropped sharp to the border of the character in order to standardize the sub-images. The image standardization is done by finding the maximum row and column with 1s and with the peak point, increase and decrease the counter until meeting the white space, or the line with all 0s. This technique is shown in figure below where a character "S" is being cropped and resized.

The image pre-processing is then followed by the image resize again to meet the network input requirement, 5 by 7 matrices, where the value of 1 will be assign to all pixel where all 10 by 10 box are filled with 1s, as shown below (Fig 13):

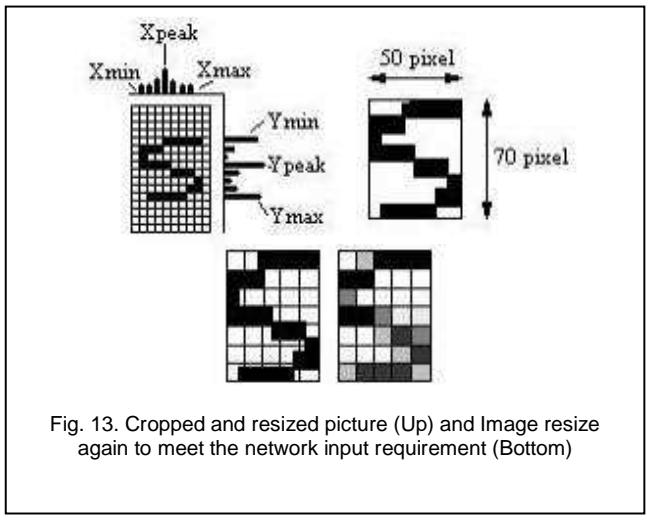

Fig. 13. Cropped and resized picture (Up) and Image resize again to meet the network input requirement (Bottom)

Finally, the 5 by 7 matrices is concatenated into a stream so that it can be feed into network 35 input neurons. The input of the network is actually the negative image of the figure, where the input range is 0 to 1, with 0 equal to black and 1 indicate white, while the value in between show the intensity of the relevant pixel. However, before resize the sub-images, another process must be gone through to eliminate the white space in the boxes. It was done using some codes in a .m file.

**Creating vectors data for the Neural Network (Objects):**

**Source Code :**

P = out(:,1:40); %input of the network

T = [eye(10) eye(10) eye(10) eye(10)];

%target of the network (Identity Matrix)

Ptest = out(:,41:50); %testing of the network

The front 4 rows will be used to train the network, while the last row will be used to evaluate the performance of the network (The Last row is for testing where the cryptography message was hidden, 'The first 6 Alphabets are 'RORVWO'. To mark this R was underline by red color, O by Green, V by Orange, W by Brown) shown in Figure. 14 and 15,16.

**Decoding from the image :** After running the neural network simulation an array of numbers were found. It was stored in an array (b). From the last row of the original figure (Hill.bmp) we can see that it was like :

R O R V W O R O V W

If we look closer than we could find that the first 6 letters were the actual message. The last row contains some other four letters just to show that the neural network was actually able to find them out too.

The vector b had 10 values : 3 2 3 5 6 2 3 2 5 6.

```
traingdx-calcgrad, Epoch 115/1000, MSE 0.0993897/0.1, Gradient 0.0238812/1e-006
traingdx, Performance goal met.

   3   2   3   5   6   2   3   2   5   6

b =

   3   2   3   5   6   2   3   2   5   6

>> Q = ['ABCDEFGHIJKLMNOPQRSTUVWXYZ'];
if b(1) == b(3)
g = Q(18);
end
 if b(2) == b(6)
g1 = Q(15);
end
if b(4) ~= b(5)
g2 = Q(22);
g3 = Q(23);
end
k = [g,g1,g,g2,g3,g1]

k =

RORVWO
```

Fig. 14. MATLAB command window of the Neural network simulation showing the vector b. and the source code of extracting the hidden message from the picture (Below)

```
>> if k(1) == Q(18)
A(1) = 18;
end
 if k(2) == Q(15)
A(2) = 15;
end
if k(3) == Q(18)
A(3) = 18;
end
 if k(4) == Q(22)
A(4) = 22;
end
if k(5) == Q(23)
A(5) = 23;
end
 if k(6) == Q(15)
A(6) = 15;
end
A = [A(1) A(3) A(5) ; A(2) A(4) A(6)];
E = [ 1 2 ; 0 3];
format rat;
B = inv(E);
C = (B*3);
D = round(mod(C*9,26));
F = mod(D*(A(1:2)'),26);
G = mod(D*(A(3:4)'),26);
H = mod(D*(A(5:6)'),26);
I = [F G H];
```

Fig. 15. Extracting the hidden code (step 1)

So the program was able to find R (Equivalent to 3), O (Equivalent to 2), V (Equivalent to 5), W (Equivalent to 6). Now using some tricky codes the hidden message and actual were revealed.

```
Q = ['ABCDEFGHIJKLMNOPQRSTUVWXYZ'];
if I(1) == 8
j = Q(8);
end
 if I(2) == 5
j1 = Q(5);
end
if I(3) == 12
j2 = Q(12);
end
 if I(4) == 16
j3 = Q(16);
end
if I(5) == 13
j4 = Q(13);
end
 if I(6) == 5
j5 = Q(5);
end
[j,j1,j2,j3,j4,j5]

ans =

HELPME
```

Fig. 16. Extracting the hidden code (step 2)

## 6  EXPERIMENTAL RESULT

The code above was successful and revealed the actual message which was encoded using Hill cipher. The neural network which performed the simulation was a Feed-Forward network with Traingdx learning algorithm and the error function was MSE. It was the best network for this kind of recognition problem. Some other networks were also used but some of them were slower and some of them were faster but unsuccessful in learning. The table 3 shows complete result.

TABLE 3
PERFORMANCE OF NEURAL NETWORKS

| Neural Network | Epoch & Err. Func | Goal = 0.1 | Comments |
|---|---|---|---|
| Feed-Forward with Traingdx | 115 Epochs & MSE | Goal met; Performance is 0.0993897 | Recognized Alphabets almost every time |
| Feed-Forward with Traingdm | 2924 Epochs & MSE | Goal met; Performance is 0.0999938 | Recognized Alphabets almost every time but slower |
| Feed-Forward with Trainlm | 3 Epochs & MSE | Goal met; Performance is 0.0808808 | Couldn't Recognize Alphabets correctly |

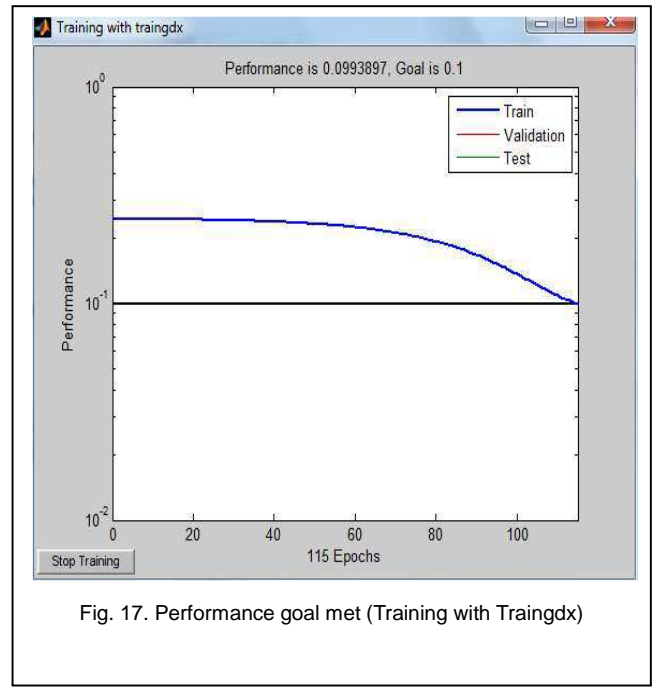

Fig. 17. Performance goal met (Training with Traingdx)

The above Figure (Fig. 17) shows the training result and after 115 epochs the training ended and the neural network successfully learned the patter in the .bmp image. There were some errors. Using the technique described in this paper it is not possible to efficiently detect handwritten alphabets or handwriting because this technique focuses darker and distinct objects on paper or in an image shown in Figure. 18.

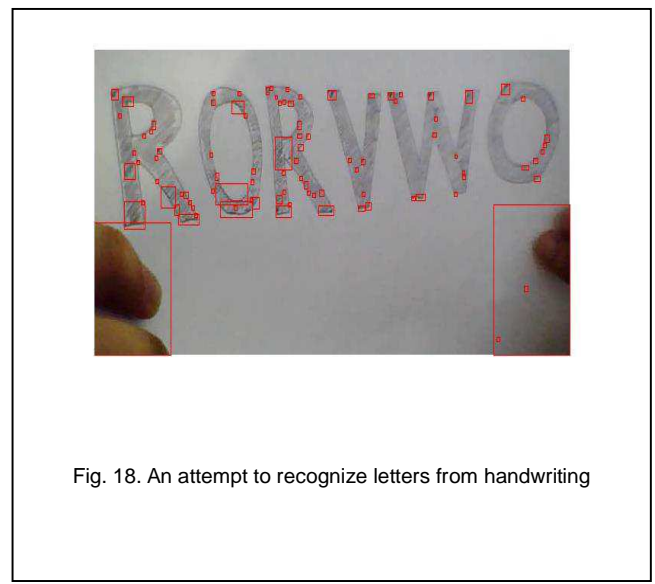

Fig. 18. An attempt to recognize letters from handwriting

In the following section some methods are discussed to recover this type of problem.

# 7 CONCLUSION AND FUTURE WORKS

The method used in this paper could be useful in various applications even in machine learning and intelligent robotics[12] although the research didn't focus on Real-Time Cryptography, rather it was based on images. Processing real time images in MATLAB is commonly known as "soft real time" since the rate at which the Image Acquisition Toolbox processes images depends on the computer microprocessor speed, the complexity of the processing algorithm, and the frame rate of the camera.

If someone wants to take this technique even further then there are some methods to do it. The Image Acquisition Toolbox uses a memory buffer to store each frame as it comes from the camera. Each frame can be acquired from the cam, loaded into MATLAB and analyzed by Image Processing Toolbox, and also processed using a set of different MATLAB commands.

In recent years, the components to construct computer vision systems have also become affordable and widely available. Video hardware like firewire cards, and television framegrabbers can have a positive impact on computer vision instruction and image processing research. Such components could facilitate lessons on image processing, visual servoing, and sensor fusion [13].

Some more technical improvements are possible in future also. Like it is common practice to divide a digital system into software and hardware components. The greatest functionality and performance occurs when functions are placed in dedicated system hardware that is highly parallel and is optimized to perform the intended operations. More often than not it is far too expensive in time and money to create such dedicated hardware so the available hardware resources are used to implement a conventional processor which executes programmable instructions in a highly sequential manner. So a dedicated system could be possible to create to detect handwritings in different environments and also in Real-Time. Some applications that could greatly benefit from this technology include Real time processing, Data encryption/decryption , RSA cryptography, Data Compression, Image and Video processing [14].

The experiment conducted in this study reveals that neural networks approaches to character recognition was simple and achieved better results. Although in the cryptography technique discussed above was simple. It was done to show that it is possible to create an efficient method using simple ideas. A more improved technique could be employed to get better result in future.

## BIOGRAPHIES

**Yousuf Ibrahim Khan** is currently pursuing his postgraduate study in Mathematics at Ahsanullah University of Science & Technology (AUST). He completed his B.Sc from American International University-Bangladesh (AIUB) under the department of Electrical and Electronic Engineering (EEE). His research interests include Artificial, Computational Intelligence, Machine vision, Image Processing, Robotics and others. He can be reached at imran_khan_eee@live.com

**Saad Mahmud Sonyy** completed his B.Sc from AIUB under the EEE department.

**S.M. Musfequr Rahman** is currently pursuing postgraduation in Mathematics at AUST. He completed his B.Sc from AIUB under the EEE department.